\documentclass[12pt]{article}
\usepackage{amsmath}
\usepackage{amsfonts}   % if you want the fonts
\usepackage{amssymb}    % if you want extra symbols
\usepackage{graphicx}   % need for figures
\usepackage{xcolor}
\usepackage{bm}
\usepackage{secdot}
\usepackage{mathptmx}
\usepackage{float}
\usepackage[utf8]{inputenc}
\usepackage{textcomp}
\usepackage[hang,flushmargin,bottom]{footmisc} % footnote format

\usepackage{titlesec}
\titleformat{\section}{\normalsize\bfseries}{\thesection.}{1em}{}	% required for heading numbering style
\titleformat*{\subsection}{\normalsize\bfseries}

\usepackage{tocloft}	% change typeset, titles, and format list of appendices/figures/tables

\usepackage{enumitem}         % to control spacing between bullets/numbered lists

\usepackage[numbers,sort&compress]{natbib} % format bibliography

\usepackage[hyphens,obeyspaces]{url}
\usepackage[hidelinks]{hyperref}
\hypersetup{
colorlinks = true,
urlcolor ={blue},
citecolor = {.},
linkcolor = {.},
anchorcolor = {.},
filecolor = {.},
menucolor = {.},
runcolor = {.}
pdftitle={},
pdfsubject={},
pdfauthor={},
pdfkeywords={}
}
\usepackage{epstopdf} % converting EPS figure files to PDF

\usepackage{fancyhdr, lastpage}	% formatting document, calculating number of pages, formatting headers
\usepackage{caption} % required for Figure labels
\usepackage{listings}
\newcommand{\code}[1]{\lstinline|#1|}

\newcommand{\subsubsubsection}[1]{\vspace{2ex}\textbf{#1}}

\usepackage{makecell}

\usepackage{pdflscape}
\setlist{nosep}

\newcommand{\pubnumber}{8397}
\newcommand{\DOI}{https://doi.org/10.6028/NIST.IR.8397}
\newcommand{\monthyear}{July 2021}
\newcommand{\paperTitle}{Guidelines on Minimum Standards for Developer Verification of Software}

\begin{document}
	\urlstyle{rm} % Format style of \url

%%%%%%%%%%%%%%%%%%%%%%%%%%%%%%%%%%%%%%%%%%%%%%%%%%%%%%%%%%%%%%%%%%%%
%   Cover Page is REQUIRED and must contain the information
%	displayed here, at a minimum. Additional artwork may be included
%	(e.g., official project/conference logo, etc.).
%	Pub Number automated based on metadata
%%%%%%%%%%%%%%%%%%%%%%%%%%%%%%%%%%%%%%%%%%%%%%%%%%%%%%%%%%%%%%%%%%%%
\begin{titlepage}
\begin{flushright}
%%%%%%%%%%%%%%%%%%%%%%%%%%%%%%%%%%%%%%%%%%%%%%%%%%%%%%%%%%%%%%%%%%%%
% 	Automated based on metadata - delete if not applicable
%%%%%%%%%%%%%%%%%%%%%%%%%%%%%%%%%%%%%%%%%%%%%%%%%%%%%%%%%%%%%%%%%%%%
\LARGE{\textbf{NISTIR \pubnumber}}\\
\vfill
%%%%%%%%%%%%%%%%%%%%%%%%%%%%%%%%%%%%%%%%%%%%%%%%%%%%%%%%%%%%%%%%%%%%
%	Title
%%%%%%%%%%%%%%%%%%%%%%%%%%%%%%%%%%%%%%%%%%%%%%%%%%%%%%%%%%%%%%%%%%%%
\Huge{\textbf{\paperTitle}}\\
\vfill
%%%%%%%%%%%%%%%%%%%%%%%%%%%%%%%%%%%%%%%%%%%%%%%%%%%%%%%%%%%%%%%%%%%%
%	Authors - add complete list of authors, affiliations will be
%   added on title page
%%%%%%%%%%%%%%%%%%%%%%%%%%%%%%%%%%%%%%%%%%%%%%%%%%%%%%%%%%%%%%%%%%%%
\large Paul E. Black\\
\large Barbara Guttman\\
\large Vadim Okun\\
\vfill
%%%%%%%%%%%%%%%%%%%%%%%%%%%%%%%%%%%%%%%%%%%%%%%%%%%%%%%%%%%%%%%%%%%%
%	The DOI is automated based on metadata.
%%%%%%%%%%%%%%%%%%%%%%%%%%%%%%%%%%%%%%%%%%%%%%%%%%%%%%%%%%%%%%%%%%%%
\normalsize This publication is available free of charge from:\\
\DOI\\
\vfill
%%%%%%%%%%%%%%%%%%%%%%%%%%%%%%%%%%%%%%%%%%%%%%%%%%%%%%%%%%%%%%%%%%%%
%	NIST LOGO - keep as-is
%%%%%%%%%%%%%%%%%%%%%%%%%%%%%%%%%%%%%%%%%%%%%%%%%%%%%%%%%%%%%%%%%%%%

\includegraphics[width=0.3\linewidth]{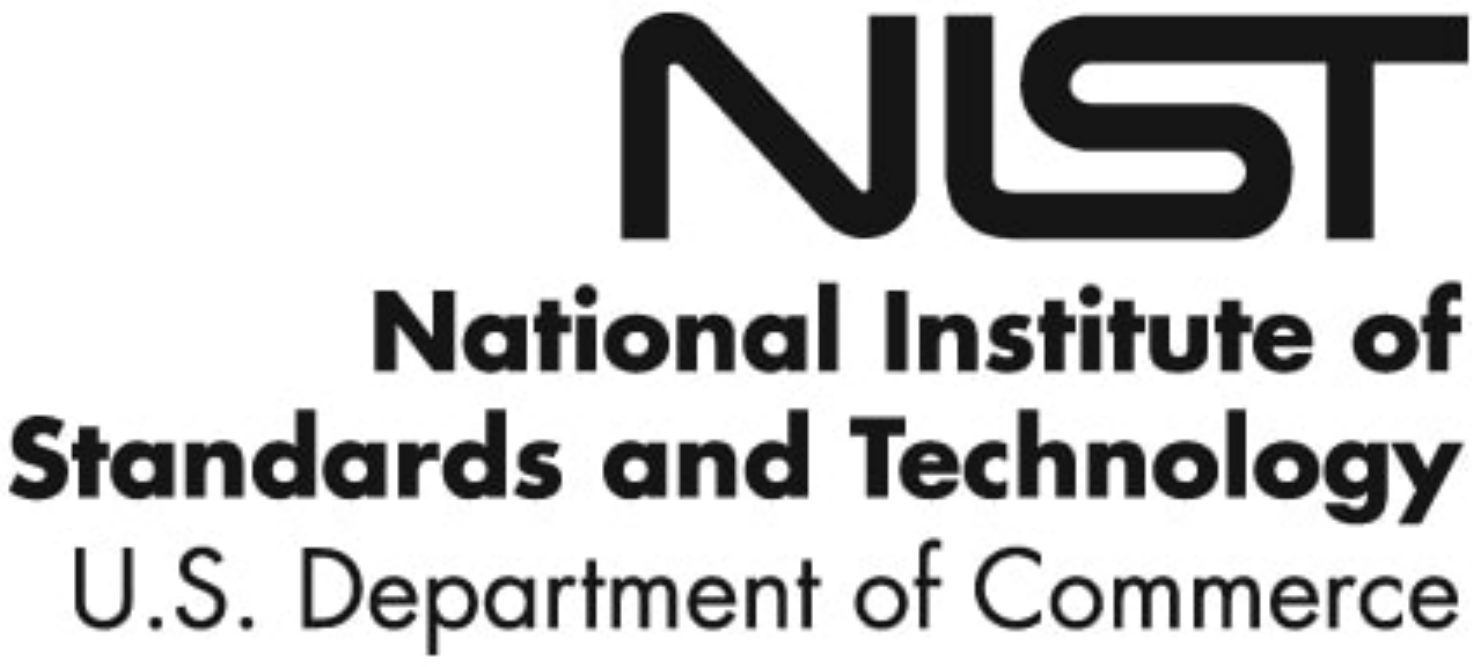}\\

\end{flushright}
\end{titlepage}
\begin{titlepage}
%%%%%%%%%%%%%%%%%%%%%%%%%%%%%%%%%%%%%%%%%%%%%%%%%%%%%%%%%%%%%%%%%%%%
%	Title Page is REQUIRED
%%%%%%%%%%%%%%%%%%%%%%%%%%%%%%%%%%%%%%%%%%%%%%%%%%%%%%%%%%%%%%%%%%%%
\begin{flushright}
%%%%%%%%%%%%%%%%%%%%%%%%%%%%%%%%%%%%%%%%%%%%%%%%%%%%%%%%%%%%%%%%%%%%
%   Publication Series & Number - automated
%%%%%%%%%%%%%%%%%%%%%%%%%%%%%%%%%%%%%%%%%%%%%%%%%%%%%%%%%%%%%%%%%%%%
\LARGE{\textbf{NISTIR \pubnumber}}\\
\vfill
%%%%%%%%%%%%%%%%%%%%%%%%%%%%%%%%%%%%%%%%%%%%%%%%%%%%%%%%%%%%%%%%%%%%
%	Title
%%%%%%%%%%%%%%%%%%%%%%%%%%%%%%%%%%%%%%%%%%%%%%%%%%%%%%%%%%%%%%%%%%%%
\Huge{\textbf{\paperTitle}}\\
\vfill
%%%%%%%%%%%%%%%%%%%%%%%%%%%%%%%%%%%%%%%%%%%%%%%%%%%%%%%%%%%%%%%%%%%%
%	Author Order and Grouping. Always identify the primary author/creator
% first (s/he does not have to be a NIST author). For publications with multiple
% authors, group authors by their organizational affiliation. The organizational
% groupings and the names within each grouping should generally be ordered by
% decreasing level of contribution.
%	For non-NIST authors, list their city and state below their organization
% name.
%	For NIST authors, include the Division and Laboratory names (but do not
% include their city and state).
%%%%%%%%%%%%%%%%%%%%%%%%%%%%%%%%%%%%%%%%%%%%%%%%%%%%%%%%%%%%%%%%%%%%
    \normalsize Paul E. Black\\
    \normalsize Barbara Guttman\\
    \normalsize Vadim Okun\\
     \textit{Software and Systems Division}\\
     \textit{Information Technology Laboratory}\\
\vfill
%%%%%%%%%%%%%%%%%%%%%%%%%%%%%%%%%%%%%%%%%%%%%%%%%%%%%%%%%%%%%%%%%%%%
%   DOI Statement - automated
%%%%%%%%%%%%%%%%%%%%%%%%%%%%%%%%%%%%%%%%%%%%%%%%%%%%%%%%%%%%%%%%%%%%
\normalsize This publication is available free of charge from:\\
\DOI\\
\vfill
%%%%%%%%%%%%%%%%%%%%%%%%%%%%%%%%%%%%%%%%%%%%%%%%%%%%%%%%%%%%%%%%%%%%
%   Date - Month and Year - automated
%%%%%%%%%%%%%%%%%%%%%%%%%%%%%%%%%%%%%%%%%%%%%%%%%%%%%%%%%%%%%%%%%%%%
\normalsize \monthyear
\vfill
%%%%%%%%%%%%%%%%%%%%%%%%%%%%%%%%%%%%%%%%%%%%%%%%%%%%%%%%%%%%%%%%%%%%
%  Department of Commerce LOGO - leave as-is
%%%%%%%%%%%%%%%%%%%%%%%%%%%%%%%%%%%%%%%%%%%%%%%%%%%%%%%%%%%%%%%%%%%%

\includegraphics[width=0.18\linewidth]{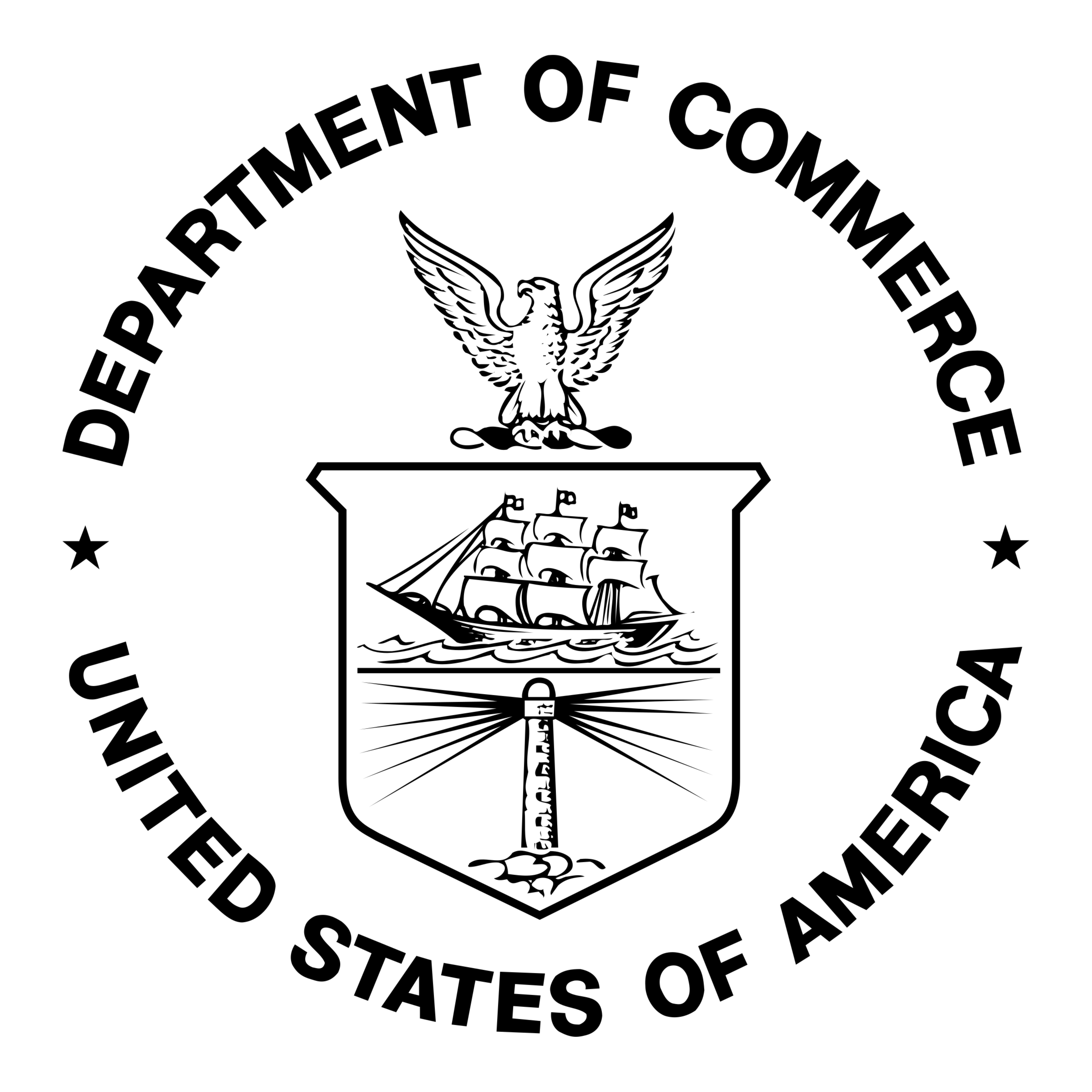}\\
 \vfill
%%%%%%%%%%%%%%%%%%%%%%%%%%%%%%%%%%%%%%%%%%%%%%%%%%%%%%%%%%%%%%%%%%%%
%  Department of Commerce & NIST Leadership
%	will be updated as changes occur
%%%%%%%%%%%%%%%%%%%%%%%%%%%%%%%%%%%%%%%%%%%%%%%%%%%%%%%%%%%%%%%%%%%%
\footnotesize U.S. Department of Commerce\\
\textit{Gina M. Raimondo., Secretary}\\
\vspace{10pt}
National Institute of Standards and Technology\\
\textit{Janes K. Olthoff, Acting NIST Director and Undersecretary of Commerce for Standards and Technology}
\end{flushright}
\end{titlepage}

\begin{titlepage}
%%%%%%%%%%%%%%%%%%%%%%%%%%%%%%%%%%%%%%%%%%%%%%%%%%%%%%%%%%%%%%%%%%%%
%   Disclaimer/CODEN page - required
%%%%%%%%%%%%%%%%%%%%%%%%%%%%%%%%%%%%%%%%%%%%%%%%%%%%%%%%%%%%%%%%%%%%
\begin{flushright}
\footnotesize
Certain commercial entities, equipment, or materials may be identified in this
document in order to describe an experimental procedure or concept adequately. Such
identification is not intended to imply recommendation or endorsement by the National
Institute of Standards and Technology, nor is it intended to imply that the entities,
materials, or equipment are necessarily the best available for the purpose.\\

\vfill
%%%%%%%%%%%%%%%%%%%%%%%%%%%%%%%%%%%%%%%%%%%%%%%%%%%%%%%%%%%%%%%%%%%%
%   This section automated - do not change
%%%%%%%%%%%%%%%%%%%%%%%%%%%%%%%%%%%%%%%%%%%%%%%%%%%%%%%%%%%%%%%%%%%%
\normalsize \textbf{National Institute of Standards and Technology \\ Interagency Report \pubnumber\\
Natl. Inst. Stand. Technol. Interag. Rep. \pubnumber,
33 %\pageref{LastPage}
pages (\monthyear)} \\
\vspace{12pt}
\textbf{This publication is available free of charge from: \DOI}
\vfill
\end{flushright}
\end{titlepage}

%%%%%%%%%%%%%%%%%%%%%%%%%%%%%%%%%%%%%%%%%%%%%%%%%%%%%%%%%%%%%%%%%%%%
%   Start front matter - page number starts with "i"
%%%%%%%%%%%%%%%%%%%%%%%%%%%%%%%%%%%%%%%%%%%%%%%%%%%%%%%%%%%%%%%%%%%%
\pagenumbering{roman}
\section*{Abstract}
\normalsize
Executive Order (EO) 14028, \emph{Improving the Nation's Cybersecurity}, 12 May 2021,
directs the National Institute of Standards and Technology (NIST) to recommend
minimum standards for software testing within 60 days.  This document describes
eleven recommendations for software verification techniques as well as providing
supplemental information about the techniques and references for further information.
It recommends the following techniques:
\begin{itemize}
\item Threat modeling to look for design-level security issues
\item Automated testing for consistency and to minimize human effort
\item Static code scanning to look for top bugs
\item Heuristic tools to look for possible hardcoded secrets
\item Use of built-in checks and protections
\item ``Black box'' test cases
\item Code-based structural test cases
\item Historical test cases
\item Fuzzing
\item Web app scanners, if applicable
\item Address included code (libraries, packages, services)
\end{itemize}

The document does not address the totality of software verification, but instead,
recommends techniques that are broadly applicable and form the minimum standards.

The document was developed by NIST in consultation with the National Security
Agen\-cy (NSA).
Additionally, we received input from numerous outside organizations through papers
submitted to a NIST workshop on the Executive Order held in early June 2021,
discussion at the workshop, as well as follow up with several of the submitters.

\vfill

{\bfseries Keywords}

\normalsize software assurance; verification; testing; static analysis; fuzzing;
code review; software security.

\vfill

{\bfseries Additional Information}

For additional information on NIST's Cybersecurity programs, projects, and
publications, visit the
\href{https://csrc.nist.gov/}{Computer Security Resource Center}.
Information on other efforts at \href{https://nist.gov/}{NIST} and in the
\href{https://nist.gov/itl/}{Information Technology Laboratory} (ITL) is also available.

% push next part to the bottom of the page
\vfill

This document was written at the National Institute of Standards and Technology by
employees of the Federal Government in the course of their official duties.  Pursuant
to Title 17, Section 105 of the United States Code, this is not subject to copyright
protection and is in the public domain.

We would appreciate acknowledgment if this document is used.

\pagebreak

{\bfseries Acknowledgments}

The authors particularly thank Fay Saydjari for catalyzing our discussion of scope;
Virginia Laurenzano for infusing DevOps Research and Assessments (DORA) principles
into the report and other material;
Larry Wagoner for numerous contributions and comments;
Steve Lipner for reviews and suggestions;
David A. Wheeler for extensive corrections and recommendations; and
Aurelien M. Delaitre, % BSIMM 11 review; Sec. 3.5 Fuzzing
William Curt Barker,
Murugiah Souppaya,
Karen Scarfone, and
Jim Lyle for their many efforts.

We thank the following for reviewing various codes, standards, guides, and other
material:
Jessica Fitzgerald-McKay, % Landwehr Building Code
Hialo Muniz, % CISQ
%Vadim Okun, % SOAR
and Yann Prono. % UL Cybersecurity Assurance Program
For the acronyms, glossary, and other content, we thank Matthew B. Lanigan,
Nhan L. Vo,
William C. Totten, and Keith W. Beatty.

We thank all those who submitted position papers applicable for
our area to our June 2021 workshop.

This document benefited greatly from additional women and men who shared their
insights and expertise during weekly conference calls between NIST and
National Security Agency (NSA) staff:
Andrew White,
Anne West,
Brad Martin,
Carol A. Lee,
Eric Mosher,
%Fay Saydjari,
Frank Taylor,
George Huber,
Jacob DePriest,
Joseph Dotzel,
%Larry Wagoner,
Michaela Bernardo,
Philip Scherer,
Ryan Martin,
Sara Hlavaty,
and
Sean Weaver.
%Virginia Laurenzano,
%Matthew B. Lanigan,
%Nhan L. Vo,
%William C. Totten,
%Keith W. Beatty,
%Jessica Fitzgerald-McKay,
%Barbara Guttman,
%Murugiah Souppaya,
%William Curt Barker,
Kevin Stine, NIST, also participated.

We also appreciate contributions from
Walter Houser.

\vfill

{\bfseries Trademark Information}

All registered trademarks or trademarks belong to their respective organizations.

\pagebreak

%%%%%%%%%%%%%%%%%%%%%%%%%%%%%%%%%%%%%%%%%%%%%%%%%%%%%%%%%%%%%%%%%%%%
%   Table of Contents is required
% 	List of Tables & Figures required if more than 5 tables/figures
%%%%%%%%%%%%%%%%%%%%%%%%%%%%%%%%%%%%%%%%%%%%%%%%%%%%%%%%%%%%%%%%%%%%
\begin{center}
  \tableofcontents
%  \listoftables
%  \listoffigures
\end{center}

\pagebreak

%%%%%%%%%%%%%%%%%%%%%%%%%%%%%%%%%%%%%%%%%%%%%%%%%%%%%%%%%%%%%%%%%%%%
%	Errata page
%%%%%%%%%%%%%%%%%%%%%%%%%%%%%%%%%%%%%%%%%%%%%%%%%%%%%%%%%%%%%%%%%%%%
\section*{Errata}

In October 2021, we made many grammatical changes due to internal paperwork to obtain
a Digital Object Identifier (DOI).  While making those, we took the opportunity to
also improve or correct text related to Interactive Application Security Testing
(IAST) and update the name of an example tool.

%\section*{} % make a visual break and don't indent next paragraph

\pagebreak

%%%%%%%%%%%%%%%%%%%%%%%%%%%%%%%%%%%%%%%%%%%%%%%%%%%%%%%%%%%%%%%%%%%%
%   Start body of text - page number starts with "1"
%%%%%%%%%%%%%%%%%%%%%%%%%%%%%%%%%%%%%%%%%%%%%%%%%%%%%%%%%%%%%%%%%%%%
\pagenumbering{arabic}

%%%%%%%%%%%%%%%%%%%%%%%%%%%%%%%%%%%%%%%%%%%%%%%%%%%%%%%%%%%%%%%%%%%%
% When referring to references in the text parenthetically, use
% the form "[1]." For example, "As Jones and Smith have shown [1];"
% However, when a reference is referred to non-parenthetically, use
% the form ". . . Ref. [1] . . ." (except at the beginning of a
% sentence where "Reference [1] . . ." is the correct form).
%%%%%%%%%%%%%%%%%%%%%%%%%%%%%%%%%%%%%%%%%%%%%%%%%%%%%%%%%%%%%%%%%%%%

%%%%%%%%%%%%%%%%%%%%%%%%%%%%%%%%%%%%%%%%%%%%%%%%%%%%%%%%%%%%%%%%%%%%
% Section references are "Sec. X".
% 	"Section X" is used at beginning of sentence.
%%%%%%%%%%%%%%%%%%%%%%%%%%%%%%%%%%%%%%%%%%%%%%%%%%%%%%%%%%%%%%%%%%%%

\section{Introduction}
\label{sec:intro}

\subsection{Overview}

To ensure that software is sufficiently safe and secure, software must be designed,
built, delivered, and maintained well.  Frequent and thorough verification by
developers
as early as possible in the software development life cycle (SDLC) is one critical
element of software security assurance.
At its highest conceptual level, we may view verification as
\emph{a mental discipline} to
increase software quality~\cite[p.~10]{AmmannOffuttSWTesting2017}.
As NIST's Secure Software Development Framework (SSDF) says, verification is used
``to identify vulnerabilities and verify compliance with security
requirements''~\cite[PW.7 and PW.8]{dodsonEtAlSSDF2020}.
According to International Organization for Standardization (ISO)/
International Electrotechnical Commission (IEC)/
Institute of Electrical and Electronics Engineers (IEEE)
12207:2017~\cite[3.1.72]{ISOIECIEE12207:2017} verification,
which is sometimes informally called ``testing'', encompasses many static and
active assurance techniques, tools, and
related processes.  They must be
employed alongside other methods to ensure a high-level of software quality.

This document recommends minimum standards of software verification by software
producers.
No single software security verification standard can encompass all types of software
and be both
specific and prescriptive while supporting efficient and effective verification.
Thus, this
document recommends guidelines for software producers to use in creating their own
processes.  To be most effective, the process must be very specific and tailored
to the software products, technology (e.g., language and platform), toolchain, and
development lifecycle model.
For information about how verification fits into the larger software
development process, see NIST's Secure Software Development Framework
(SSDF)~\cite{dodsonEtAlSSDF2020}.

\subsection{Charge}

This document is a response to the 12 May 2021 Executive Order (EO) 14028 on
Improving
the Nation's Cybersecurity~\cite{EOImproveCybersecurity2021}.
This document responds to Sec.~4. Enhancing Software Supply Chain Security,
subsection (r):

\vspace{1ex}

\hspace*{.5em}
\begin{minipage}{.9\textwidth}
``\ldots guidelines recommending minimum standards for vendors' testing of their
software source code,
including identifying recommended types of manual or automated testing (such as code
review tools, static and dynamic analysis, software composition tools, and
penetration testing).'' \cite[4(r)]{EOImproveCybersecurity2021}
\end{minipage}

\subsection{Scope}
\label{sec:scope}

This section clarifies or interprets terms that form the basis for the scope of this
document.

We define ``software'' as executable computer programs.

We exclude from our scope ancillary yet vital
material such as configuration files,
file or execution permissions, operational procedures, and hardware.

\vspace{1ex}

Many kinds of software require specialized testing regimes in addition to the minimum
standards recommended in Sec.~\ref{sec:recMinStandardTesting}.
For example, real-time software, firmware (microcode),
embedded/cyberphysical software, distributed algorithms,
machine learning (ML) or neural net code,
% Barbara called it operational technology (OT).  Kevin Stein said SP 800-82 r3
% will standardize on the term "control system" as that is what practitioners use.
control systems, % building, water, pipeline, power, etc., control systems
mobile applications, safety-critical systems, and
cryptographic software.  We do not address this specialized testing further.
We do suggest minimum testing techniques to use for software that is connected to
a network and
parallel/multi-threaded software.

As a special note, testing requirements for safety-critical systems are
addressed by
their respective regulatory agencies.

\vspace{1em}

While the EO uses the term ``software source code'', the intent is
much broader and includes software in general including binaries, bytecode, and
executables, such as
libraries and packages.
We acknowledge that it is not possible to examine these as thoroughly
and efficiently as human-readable source code.

We exclude from consideration here the verification or validation of security
functional requirements and specifications, except as references for testing.

\vspace{1em}

We understand the informal term ``testing'' as any technique or procedure performed
on the software itself
to gain assurance that the software will perform as desired, has the
necessary properties, and has no important
vulnerabilities.  We use the ISO/IEC/IEEE term ``verification'' instead.
Verification includes methods such as static analysis and code review, in addition to
dynamic analysis or running programs (``testing'' in a narrower sense).

We exclude from our treatment of verification other key elements of
software development that
contribute to software assurance, such as programmer training,
expertise, or certification, evidence
from prior or subsequent software products, process,
correct-by-construction or model-based
methods, supply chain and compilation assurance techniques, and
failures reported during operational use.

Verification assumes standard language semantics, correct and
robust compilation or interpretation engines, and a reliable and accurate
execution environment, such as containers, virtual machines, operating
systems, and hardware.  Verification may or may not be performed in the intended
operational environment.

\vspace{1ex}

Note that verification must be based on some references, such as the software
specifications, coding standards (e.g., Motor Industry Software Reliability
Association (MISRA) C~\cite{MISRAC2012}), collections of properties, security
policies, or lists of common weaknesses.

\vspace{1em}

While the EO uses the term ``vendors' testing'', the intent is much
broader and includes developers as well.  A developer and a vendor may be the same
entity, but many vendors include software from outside sources.  A software vendor
may redo verification on
software packages developed by other entities.
Although the EO mentions commercial
software~\cite[Sec.~4(a)]{EOImproveCybersecurity2021}, this guideline is written for
all software developers, including those employed by the government and developers of
open-source software (OSS).
The techniques and procedures presented in this document might be used by software
developers to verify reused software that they incorporate in their product,
customers acquiring software,
entities accepting contracted software, or a third-party lab. However, these
are not the intended audience of this document since this assurance effort should be
applied as early in the development process as possible.

\vspace{1em}

This document presents ``minimum standards''. That is, this document is not a guide
to most effective practices or
recommended practices.  Instead, its purposes are to (1) set a lower bar for software
verification by indicating techniques that developers should have already been using
and (2) serve as a basis for
mandated standards in the future.

\subsection{How Aspects of Verification Relate}
\label{sec:testingAspects}

\begin{figure}[ht]
  \centerline{\includegraphics[width=0.8\linewidth]{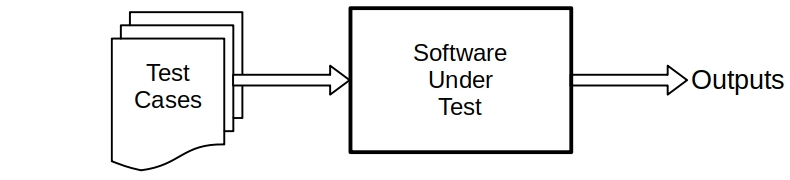}}
  \caption{The basic dynamic testing process is to deliver a set of test cases to the
    software being tested and examine the outputs.}
  \label{fig:basic testing}
\end{figure}

This section explains how code-based analysis and reviews relate to dynamic
analysis.
The fundamental process of dynamic testing of software is shown in
Fig.~\ref{fig:basic testing}.  By the time the target software has reached this
stage, it should have undergone static analysis by the compiler or other tools.
In dynamic testing, the software is run on many test cases,
and the outputs are examined.  One advantage of dynamic testing is that it has few,
if any, false positives.  For a general model of dynamic testing see
\cite[Sec.~3.5.1]{FMStatSW2019}, which also cites publications.

Verification must be automated in order for thousands of tests to be
accurately performed and for the results to be precisely checked.
Automation also allows verification to be efficiently repeated often.

\begin{figure}[ht]
  \centerline{\includegraphics[width=0.8\linewidth]{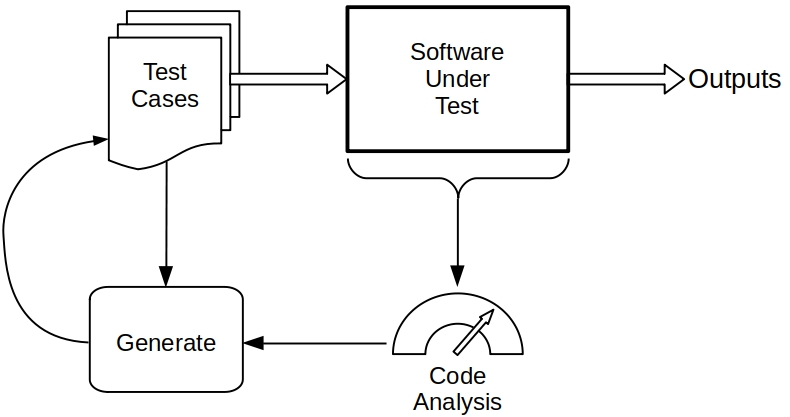}}
  \caption{A more elaborate diagram of the verification process adding how some
    test cases are generated and showing
    how code analysis fits.}
  \label{fig:testing details}
\end{figure}

Figure~\ref{fig:testing details} provides more details about the process of gaining
assurance of software.
It shows that some test cases result from a combination of the current set of test
cases and from
analysis of the code, either entirely by static consideration of the software
or by analysis of coverage during test case execution.
Sections~\ref{sec:blackBoxSources} and \ref{sec:codeBasedSources} briefly
discuss black box and code-based test cases.

Code analysis examines the code itself to check that it has desired properties,
to identify weaknesses,
and to compute metrics of test completeness.  It is also used to diagnose the
cause of faults discovered during testing.
See Sec.~\ref{sec:staticAnalysis}
for details.

This analysis can determine which statements, routines, paths, etc., were
exercised by tests and can produce measures of how complete testing was.
Code analysis can also monitor for faults such as exceptions, memory
leaks, unencrypted critical information, null pointers, SQL injection, or cross-site
scripting.

During testing, such hybrid analysis can also drive active automatic testing, see
Secs.~\ref{sec:minFuzzing} and \ref{sec:minWebAppScan}, and is used for Interactive
Application Security
Testing (IAST).  Runtime Application Self-Protection (RASP) monitors the program
during operation for internal security faults before they become system failures.
IAST and RASP may also scrutinize the output, for example, for sensitive data being
transmitted that is not encrypted.

\subsection{Document Outline}

Section~\ref{sec:recMinStandardTesting} begins with a succinct guideline
recommending minimum standard techniques for developers to use to verify their
software.  It then expands on the techniques.
Section~\ref{sec:techniqueDetails} is informative.  That is, it is not part of the
recommended minimum.  It provides background and
supplemental material about the techniques, including references, more thorough
variations and alternatives, and example tools.
Section~\ref{sec:beyondMinimum} summarizes how software must and can be built well from
the beginning.
Finally, Section~\ref{sec:sourceDocs} lists materials we consulted for this document.

\section{Recommended Minimum Standard for Developer Testing}
\label{sec:recMinStandardTesting}

Gaining assurance that software does what its developers intended and is sufficiently
free from vulnerabilities---either intentionally designed into the software or
accidentally inserted at any time during its life cycle---requires the use of many
interrelated
techniques.  % definition after NISTIR 8074 Volume 2 page 43 or NIST SP 800-163
This guideline recommends the following minimum standard for developer testing:
\begin{itemize}
  \item Do threat modeling (see Sec.~\ref{sec:threat modeling}).

\hspace*{-2.5em}
Using automated testing (Sec.~\ref{sec:automatedTesting})
for static and dynamic analysis,

  \item Do static (code-based) analysis
  \begin{itemize}
    \item Use a code scanner to look for top bugs (\ref{sec:briefStaticAnalysis}).

    \vspace{1ex}
    \item Use heuristic tools to look for hardcoded secrets and identify small
      sections of software that may warrant focused manual code reviews
      (\ref{sec:hardcodedSecrets}).
  \end{itemize}

  \item Do dynamic analysis (i.e., run the program)
  \begin{itemize}
    \item Run the program with built-in checks and protections
      (\ref{sec:languageProtection}).
    \item Create ``black box'' test cases, e.g., from specifications, input boundary
      analysis, and those motivated by threat modeling (\ref{sec:blackBoxSources}).
    \item Create code-based (structural) test cases.  Add cases as necessary to reach
      at least 80\,\%
      coverage (\ref{sec:codeBasedSources}).
    \item Use test cases that were designed to catch previous bugs
      (\ref{sec:bugSpecificTest}).

    \vspace{1ex}
    \item Run a fuzzer (\ref{sec:minFuzzing}).  If the software runs a web service,
      run a web application scanner, too
      (\ref{sec:minWebAppScan}).
  \end{itemize}

  \vspace{1ex}
  \item Correct the ``must fix'' bugs that are uncovered and
  improve the process to prevent similar bugs in the future, or at least
  catch them earlier~\cite[RV.3]{dodsonEtAlSSDF2020}.

  \item Use similar techniques to gain assurance that included libraries, packages,
  services, etc., are no less secure than the code~(\ref{sec:includedCode}).

\end{itemize}

The rest of this section provides additional information about each aspect of the
recommended minimum standard.

\subsection{Threat Modeling}
\label{sec:threat modeling}

We recommend using threat modeling early in order to identify design-level security
issues and to focus verification.
Threat-modeling methods create an abstraction of the system,
profiles of potential attackers, including their goals and methods,
and a catalog of potential threats~\cite{ShevchenkoEtAlTM2018}.
See also \cite{ShostackTM2014}.
Shevchenko et al.~\cite{ShevchenkoEtAlTM2018} lists twelve threat-modeling methods,
pointing out
that software needs should drive the method(s) used.
Threat modeling should be done \emph{multiple} times during development,
especially when developing
new capabilities, to capture new threats and improve
modeling~\cite{KeaneTM2021}.
The DoD Enterprise DevSecOps Reference Design document of August 2019 includes a
diagram of how threat modeling fits into software development (Dev), security (Sec),
and operations (Ops)~\cite[Fig. 3]{DoDDevSecOpsRef2019}.  DevSecOps is an
organizational software engineering culture and practice focused on unifying
development, security, and operations aspects related to software.

Test cases should be more comprehensive in areas of greatest consequences, as
indicated by the threat assessment or threat scenarios.
Threat modeling can also indicate which input vectors are of most concern.  Testing
variations of these particular inputs should be higher priority.
Threat modeling may reveal that certain small pieces of code, typically less than
100 lines, pose significant risk.  Such code may warrant manual code review to answer
specific questions such as, ``does the software require authorization when it
should?'' and ``do the software interfaces check and validate input?''
See Sec.~\ref{sec:propertyReview} for more on manual reviews.

\subsection{Automated Testing}
\label{sec:automatedTesting}

Automated support for verification can be as simple as a script that reruns static
analysis, then runs the program on a set of inputs, captures the outputs, and
compares the outputs to expected results.  It can be as sophisticated as a tool that
sets up the environment, runs the test, then
checks for success.  Some test tools drive the interface to a web-enabled
application, letting the tester specify high-level commands, such as ``click this
button'' or ``put the following text in a box'' instead of maneuvering a mouse
pointer to a certain place on a rendered screen and passing events.  Advanced tools
produce reports of what code passes their tests or summaries of the number of tests
passed for modules or subsystems.

We recommend automated verification to
\begin{itemize}
  \item ensure that static analysis does not report new weaknesses,
  \item run tests consistently,
  \item check results accurately, and
  \item minimize the need for human effort and expertise.
\end{itemize}
Automated verification can be integrated into the existing workflow or issue tracking
system~\cite[PO.3]{dodsonEtAlSSDF2020}.  Because verification is automated, it can be
repeated often, for instance, upon
every commit or before an issue is retired.

\subsection{Code-Based, or Static, Analysis}
\label{sec:briefStaticAnalysis}

Although there are hybrids, analysis may generally be divided into two approaches:
1) code-based
or static analysis (e.g., Static Application Security Testing---SAST) and
2) execution-based or dynamic analysis (e.g., Dynamic Application Security
Testing---DAST).
Pure code-based analysis
is independent of program execution.  A static code scanner reasons about the code as
written, in somewhat the same fashion as a human code reviewer. Questions that a
scanner may address include:
\begin{itemize}
\item Does this software \emph{always} satisfy the required security policy?
\item Does it satisfy important properties?
\item Would any input cause it to fail?
\end{itemize}

We recommend using a static analysis tool to check code for many kinds of
vulnerabilities, see Sec.~\ref{sec:noTopBugs}, and for compliance with the
organization's coding standards.
For multi-threaded or parallel processing software, use a scanner capable of
detecting race conditions.
See Sec.~\ref{sec:staticAnalysis} for example tools and more guidelines.

Static scanners range in sophistication from simply searching for any use of a
deprecated function to looking for patterns indicating possible vulnerabilities to
being able to verify that a piece of code faithfully implements a communication
protocol.  In addition to closed source tools, there are powerful free and
open-source tools that provide extensive analyst aids, such as control flows and
data values that lead to a violation.

Static source code analysis should be done as soon as code is written.  Small pieces
of code can be checked before large executable pieces are complete.

\subsection{Review for Hardcoded Secrets}
\label{sec:hardcodedSecrets}

We recommend using heuristic tools to examine the code for hardcoded passwords and
private encryption keys. Such tools are feasible since functions or services taking
these as parameters have specific interfaces.
Dynamic testing is unlikely to uncover such unwanted code.

While the primary method to reduce the chance of malicious code is integrity
measures, heuristic tools may assist by identifying small sections of code that are
suspicious, possibly triggering manual review.

Section~\ref{sec:propertyReview} lists additional properties that might be checked
during scans or reviews.

\subsection{Run with Language-Provided Checks and Protection}
\label{sec:languageProtection}

Programming languages, both compiled and interpreted, provide many built-in checks
and protections.  Use such capabilities both during development and in the software
shipped~\cite[PW.6.2]{dodsonEtAlSSDF2020}.
Enable hardware and operating system security and vulnerability mitigation
mechanisms, too (see Sec.~\ref{sec:compileFlags}).

For software written in languages that are not memory-safe, consider using techniques
that enforce memory safety (see Sec.~\ref{sec:memorySafeCompilation}).

Interpreted languages typically have significant security enforcement built-in,
although additional measures can be enabled.  In addition, you may
use a static analyzer, sometimes called a ``linter'', which checks for dangerous
functions, problematic parameters, and other possible vulnerabilities
(see Sec.~\ref{sec:staticAnalysis}).

\vspace{1em}

Even with these checks, programs must be executed.
Executing all possible inputs is impossible except for programs with the tiniest
input spaces.  Hence, developers must select or construct the test cases to be used.
Static code analysis can
add assurance in the gaps between test cases, but selective test execution is still
required.  Many principles can guide the choice of test cases.

\subsection{Black Box Test Cases}
\label{sec:blackBoxSources}

``Black box'' tests are not based on the implementation or the particular code.
Instead, they are based on functional specifications or requirements,
negative tests (invalid inputs and testing what the software should \emph{not}
do)~\cite[p.~8-5, Sec.~8.B]{SOAR2016},
% these are all in the same section now
%Sec.~\ref{sec:genSecPrincip}
%and Sec.~\ref{sec:riskAssessTests}
denial of service and overload, described in Sec.~\ref{sec:DOSandOverload},
input boundary analysis, and
input combinations~\cite{combinTestKuhnetal2009, combinTestKuhnetal2015}.

Tests cases should be more comprehensive in areas indicated as security sensitive or
critical by general security principles.

If you can formally prove that classes of errors cannot occur, some of the testing
described above may not be needed.  Additionally, rigorous process metrics may show
that the benefit of some testing is small compared to the cost.

\subsection{Code-Based Test Cases}
\label{sec:codeBasedSources}

Code-based, or structural, test cases are based on the implementation, that is, the
specifics of the code.  For instance, suppose the software is required to handle up to
one million items.  The programmer may decide to implement the software to handle 100
items or fewer in a statically-allocated table but dynamically allocate memory if
there are more than 100 items.  For this implementation, it is useful to have cases
with exactly 99, 100, and 101 items in order to test for bugs in switching between
approaches.  Memory alignment concerns may indicate additional tests.
These important test cases could not have been determined by only
considering the specifications.

Code-based test cases may also come from coverage metrics.  As hinted at in
Fig.~\ref{fig:testing details}, when tests are run, the software may record which
branches, blocks, function calls, etc., in the code are exercised or ``covered''.
Tools then analyze this information to compute metrics.  Additional test cases can
be added to increase coverage.

Most code should be executed during unit testing.  We recommend that executing
the test suite achieves a minimum of 80\,\% statement
coverage~\cite{CornettCoverage2013} (see Sec.~\ref{sec:blockCoverage}).

\subsection{Historical Test Cases}
\label{sec:bugSpecificTest}

Some test cases are created specifically to show the presence (and later, the
absence) of a bug.  These are sometimes called ``regression tests''.
These test cases are an important source of tests until the process is
mature enough to cover them, that is, until a ``first principles'' assurance
approach is adopted that would detect or preclude the bug.
An even better option is adoption of an assurance approach, such as
choice of language, that precludes the bug entirely.

Inputs recorded from production operations may also be
good sources of test cases.

\subsection{Fuzzing}
\label{sec:minFuzzing}

We recommend using a fuzzer, see Sec.~\ref{sec:fuzzing},
which performs automatic active testing; fuzzers create huge numbers of inputs
during testing.  Typically, only a tiny fraction of the inputs trigger code
problems.

In addition, these tools only perform a general check to determine that the software
handled the test correctly.  Typically, only broad output
characteristics and gross behavior, such as application crashes, are monitored.

The advantage of generality is that such tools can try an immense number of inputs
with minimal human supervision.  The tools can be programmed with inputs that often
reveal bugs, such as very long or empty inputs and special characters.

\subsection{Web Application Scanning}
\label{sec:minWebAppScan}

If the software provides a web service, use a
dynamic application security testing (DAST) tool,
e.g., web application scanner, see Sec.~\ref{sec:WebAppScanning},
or Interactive Application Security Testing (IAST) tool
to detect vulnerabilities.

As with fuzzers, web app scanners create inputs as they run.  A web app scanner
monitors for general unusual behavior.  A hybrid or IAST tool may also monitor
program execution for internal faults.  When an input causes some
detectable anomaly, the tool can use variations of the input to probe for failures.

\subsection{Check Included Software Components}
\label{sec:includedCode}

Use the verification techniques recommended in this section to gain assurance
that included code is at least as secure as code developed
locally~\cite[PW.3]{dodsonEtAlSSDF2020}.  Some assurance
may come from self-certification or partially self-certified information, such as
the Core Infrastructure Initiative (CII)
Best Practices badge~\cite{CIIBadge} or trusted third-party examination.

The components of the software must be continually monitored against databases of
known vulnerabilities; a new vulnerability in
existing code may be reported at any time.

A Software Composition Analysis (SCA) or Origin Analyzer (OA) tool can help you
identify what open-source
libraries, suites, packages, bundles, kits, etc., the software uses.  These
tools can aid in determining what software is really imported, identifying
reused software (including open-source software), and noting software that is
out of date or has known vulnerabilities (see Sec.~\ref{sec:noKnownVulns}).

\section{Background and Supplemental Information About Techniques}
\label{sec:techniqueDetails}

This section is informative, not part of the recommended minimum.
It provides more details about techniques and approaches.
Subsections include
information such as variations, additional cautions and considerations, example
tools, and tables of related standards, guides, or references.

\subsection{Supplemental: Built-in Language Protection}
\label{sec:compileFlags}

Programming languages have various protections built into them that preclude some
vulnerabilities, warn about poorly written or insecure code, or protect programs
during execution.  For instance, many languages are memory-safe by default.  Others
only have flags and options to activate their protections.  All such protections
should be used as much as possible~\cite[PW.6.2]{dodsonEtAlSSDF2020}.

For instance, gcc has flags that enable
\begin{itemize}
  \item run-time buffer overflow detection, % -D\_FORTIFY\_SOURCE=2
  \item run-time bounds checking for C++ strings and containers, %-D\_GLIBCXX\_ASSERTIONS
  \item address space layout randomization (ASLR), % -fPIC Create position-independent code
  \item increased reliability of stack overflow detection, % -fstack-clash-protection
  \item stack smashing protector, % -fstack-protector or -fstack-protector-all and
                          % -fstack-protector-strong
  \item control flow integrity protection, % -mcet -fcf-protection
  \item rejection of potentially unsafe format string arguments, % -Werror=format-security
  \item rejection of missing function prototypes, and % -Werror=implicit-function-declaration
  \item reporting of many other warnings and errors. % -Wall
\end{itemize}

Similarly, the Visual Studio 2019 option ``{\tt /sdl}'' enables checks comparable to
those described above for gcc.

Interpreted languages typically have significant security enforcement built-in,
although additional measures can be enabled.
As an example of an interpreted language, Perl has a ``taint'' mode, enabled by the
``{\tt -T}'' command line flag, that
``turns on various checks, such as checking path directories to make sure they aren't
writable by others.''~\cite[10.2]{WheelerSecureProgramming}.
The ``{\tt -w}'' command line option helps as do other measures explained in the
Perl security document, perlsec~\cite{perlsec}.
JavaScript has a ``use strict'' directive ``to indicate that the
code should be executed in ``strict mode''.  With strict mode, you cannot, for
example, use undeclared variables.''~\cite{JSUseStrict}

In addition, you may use a static analyzer, sometimes called a ``linter'', to check
for dangerous function or problematic parameters in interpreted languages
(see Sec.~\ref{sec:staticAnalysis}).

In addition to capabilities provided by the language itself,
%Andy White writes, (3 May 2021)
you can use hardware (HW) and operating system (OS) mechanisms to ensure control flow
integrity, for instance, Intel's Control-flow Enforcement Technology (CET) or ARM
Pointer Authentication and landing points.
There are compiler options that create opcodes so that if the software is running on
hardware, operating systems, or processes with these enabled, these mechanisms will
be invoked.
All x86 and ARM chips in production have and
will have this capability.  Most OSs now support it.

% Andy White adds, (22 June 2021)
Users should also take advantage of HW and OS mechanisms as they update technology by
ensuring the HW or OS that they are upgrading to include these HW-based features.
These mechanisms help prevent memory corruption bugs that are not detected by
verification during development from being exploited.

\begin{table}[H]
\centering
\begin{tabular}{|p{0.65\textwidth}|p{0.25\textwidth}|}
\hline
{\bfseries Technique, Principle, or Directive} &
{\bfseries Reference} \\

\hline
``Applying warning flags'' &
\cite[p.~8-4, Sec.~8.B]{SOAR2016}
\\

\hline
Using stack protection &
\cite{cacaBSI2008}
\\

\hline
Prevent execution of data memory
Principle 17 &
\cite[p.~9]{ULIoTSecPrinciples2017}
\\

\hline
\end{tabular}
\caption{Related Standards, Guides, or References for Built-in Language Protection}
\end{table}

\subsection{Supplemental: Memory-Safe Compilation}
\label{sec:memorySafeCompilation}

Some languages, such as C and C++, are not memory-safe.  A minor memory access error
can lead to vulnerabilities such as privilege escalation, denial of service, data
corruption, or exfiltration of data.

Many languages are memory-safe by default but have mechanisms to disable those
safeties when needed, e.g., for critical performance requirements.  Where practical,
use memory-safe languages and limit disabling memory safety mechanisms.

For software written languages that are not memory-safe, consider using automated source
code transformations or compiler techniques that enforce memory safety.

Requesting memory mapping to a fixed (hardcoded) address subverts address space
layout randomization (ASLR).  This should be mitigated by enabling
appropriate compile flag(s) (see Sec.~\ref{sec:compileFlags}).

\subsubsubsection{Example Tools}

Baggy Bounds Checking, CodeHawk, SoftBoundCETS, and WIT.

\begin{table}[H]
\centering
\begin{tabular}{|p{0.65\textwidth}|p{0.25\textwidth}|}
\hline
{\bfseries Technique, Principle, or Directive} &
{\bfseries Reference} \\

\hline
``Applying warning flags'' &
\cite[p.~8-4, Sec.~8.B]{SOAR2016}
\\

\hline
using stack protection &
\cite{cacaBSI2008}
\\

\hline
Element A ``avoid/detect/remove specific types of vulnerabilities at the
implementation stage'' &
\cite[p.~9--12]{medicalDevBCode2015}
\\

\hline
FPT\_AEX\_EXT.1 Anti-Exploitation Capabilities
``The application shall not request to map memory at an explicit address except for
[{\bfseries assignment}: \emph{list of explicit exceptions}].'' &
\cite{PPforAppSoftware2019v1.3}
\\

\hline
\end{tabular}
\caption{Related Standards, Guides, or References for Memory-Safe Compilation}
\end{table}

\subsection{Supplemental: Coverage Metrics}
\label{sec:blockCoverage}

Exhaustive testing is intractable for all but the simplest programs,
yet thorough testing
is necessary to reduce software vulnerabilities.  Coverage criteria are a way to
define what needs to be tested and when the testing objective is achieved.  For
instance, ``statement coverage'' measures the statements in the code that are
executed at least once, i.e., statements that are ``covered''.

Checking coverage identifies parts of code that have not been thoroughly tested
and, thus, are
more likely to have bugs.  The percentage of coverage,
e.g., 80\,\% statement coverage,
is one measure of the thoroughness of a test suite.  Test cases can be added to
exercise code or paths that were not executed.
Low coverage indicates inadequate testing, but very high code coverage guarantees
little~\cite{CornettCoverage2013}.

Statement coverage is the weakest criterion widely used.  For instance, consider an
``if'' statement with only a ``then'' branch, that is, without an ``else'' branch.
Knowing that statements in the ``then'' branch were executed does not guarantee
that any test explored what happens
when the condition is false and the body is not executed at all.  ``Branch coverage''
requires that every branch is taken.  In the absence of early exits, full branch
coverage implies full block
coverage, so it is stronger than block coverage.
Data flow and mutation are stronger coverage criteria~\cite{zhuHallMay}.

Most generally,
``\ldots all test coverage criteria can be boiled down to a few dozen
criteria on just \textbf{four} mathematical structures: input domains,
graphs, logic expressions, and syntax descriptions
(grammars).''~\cite[p.~26, Sec.~2.4]{AmmannOffuttSWTesting2017}
An example of testing based on input domains is combinatorial
testing~\cite{combinTestKuhnetal2009, combinTestKuhnetal2015}, which partitions the
input space into groups and tests all n-way combinations of the groups.
Block, branch, and data flow coverage are graph coverage criteria.  Criteria based on
logic expressions, such as Modified Condition  Decision Coverage (MCDC), require
making various truth assignments to the expressions.

Syntax description criteria are exemplified by mutation testing that deliberately and 
systematically creates software variants with small syntactic changes that are
likely to be errors.  For instance, the ``less than'' operator (\textless) might be
replaced by ``greater than or equal to'' (\textgreater =).
If a test set distinguishes the original program from each slight variation, the test
set is exercising the program adequately.  Mutation testing can be applied to
specifications as well as programs.
% Mutation is a structured form of bebugging
% Bebugging suggested by Andy White 14 May 2021
% Deliberately inject bugs at different times in the software development life cycle.
% The number reaching final production without being detected or removed can be used
% to estimate the total number of residual bugs.
% The appropriate mathematical and statistical concepts have been worked out as the
% capture-recapture method for estimating animal populations.

Note: the code may be compiled with certain flags to measure coverage, then compiled
again with different flags for shipment. There needs to be assurance that the source
and any included binaries used to build the shipped products match those verified and
measured for coverage.

\begin{table}[H]
\centering
\begin{tabular}{|p{0.65\textwidth}|p{0.25\textwidth}|}
\hline
{\bfseries Technique, Principle, or Directive} &
{\bfseries Reference} \\

\hline
``Test coverage analyzer'' &
\cite[p.~8-6, Sec.~8.B]{SOAR2016}
\\

\hline
``Relevant Metrics'' &
\cite{rbfstBSI2013}
\\

\hline
[ST3.4] ``Leverage coverage analysis'' &
\cite[p.~78]{BSIMM112020}
\\

\hline
\end{tabular}
\caption{Related Standards, Guides, or References for Coverage Metrics}
\end{table}

\subsection{Supplemental: Fuzzing}
\label{sec:fuzzing}

Fuzzers and related automated randomized test generation techniques are particularly
useful when run often during software development.  Continuous fuzzing on changing
code bases can help catch unexpected bugs early.
``Fuzz testing is effective for
finding vulnerabilities because most modern programs have extremely large input
spaces, while test coverage of that space is comparatively
small.''~\cite{OkunFongFuzzTesting2015}
Pre-release fuzzing is particularly useful, as it denies malicious parties
use of the very same tool to find bugs to exploit.

Fuzzing is a mostly automated process that may require relatively modest ongoing manual
labor.  It usually requires
a harness to feed generated inputs to the software under test.  In some case, unit test
harnesses may be used.  Fuzzing is computationally-intensive and yields best results
when performed at scale.

Fuzzing components separately can be efficient and improve code coverage.
In this case, the entire system must also be fuzzed as one to investigate whether
components work properly when used together.

One key benefit of fuzzing is that it typically produces actual positive tests
for bugs, not just
static warnings.  When a fuzzer finds a failure, the triggering input can be saved and
added to the regular test corpus.  Developers can use the execution trace leading to
the failure to understand and fix the bug.  This may not be the case when failures
are non-deterministic, for instance, in the presence of threads, multiple interacting
processes, or distributed computing.

Fuzzing approaches can be grouped into two categories based on how they create input:
mutation based and generation based.  Mutation-based fuzzing modifies
existing inputs, e.g., from
unit tests, to generate new inputs.  Generation-based fuzzing produces random inputs
from a formal grammar that describes well-formed inputs.  Using both gains the
advantages of both mutation and generation
fuzzers.  Using both approaches can cover a larger set of test case scenarios,
improve code coverage, and increase the chance of finding vulnerabilities missed by
techniques such as code reviews.

Mutation-based fuzzing is easy to set up since it needs little or no description of
the structure.  Mutations to existing inputs may be random or may follow
heuristics.  Unguided fuzzing typically shallowly explores execution paths.
For instance, completely random inputs to date fields are unlikely to be valid.
Even most random inputs that are two-digit days (DD), three-letter month
abbreviations (Mmm), and four-digit years (YYYY) will be
rejected.  Constraining days to be 1--31, months to be Jan, Feb, Mar, etc.,
and years to be within 20 years of today may still not exercise leap-century
calculations or deeper logic.
Generation-based fuzzing can pass program validation to achieve deeper testing but
typically requires far more time and expertise to set up.

Modern mutation-based fuzzers explore execution paths more deeply than unguided
fuzzers by using methods such as instrumentation and symbolic execution to take paths
that have not yet been explored.  Coverage-guided fuzzers, such as AFL++, Honggfuzz,
and libFuzzer, aim at maximizing code coverage.

To further improve effectiveness, design review, which should be first done near
the beginning of development, may indicate which input vectors
are of most concern.  Fuzzing these particular inputs should be prioritized.

Fuzzing is often used with special instrumentation to increase the likelihood of
detecting faults.  For instance, memory issues can be detected by tools such as
Address Sanitizer (ASAN) or Valgrind.  This instrumentation can cause significant
overhead but enables detection of out-of-bounds
memory access even if the fault would not cause a crash.

\subsubsubsection{Example Tools}

American Fuzzy Lop Plus Plus (AFL++), Driller, dtls-fuzzer, Eclipser, Honggfuzz,
Jazzer, libFuzzer, Mayhem, Peach, Pulsar, Radamsa, and zzuf.

\begin{table}[H]
\centering
\begin{tabular}{|p{0.65\textwidth}|p{0.25\textwidth}|}
\hline
{\bfseries Technique, Principle, or Directive} &
{\bfseries Reference} \\

\hline
PW.8: Test Executable Code to Identify Vulnerabilities and Verify Compliance with
Security Requirements &
\cite{dodsonEtAlSSDF2020}
\\

\hline
``Fuzz testing'' &
\cite[p.~8-5, Sec.~8.B]{SOAR2016}
\\

\hline
Malformed Input Testing (Fuzzing) &
\cite[slide 8]{ULandCybersecurity2011}
\\

\hline
[ST2.6] ``Perform fuzz testing customized to application API'' &
\cite[p.~78]{BSIMM112020}
\\

\hline
\end{tabular}
\caption{Related Standards, Guides, or References for Fuzzing}
\end{table}

\subsection{Supplemental: Web Application Scanning}
\label{sec:WebAppScanning}

DAST tools, such as web app scanners, test software in operation.  IAST tools check
software in operation.  Such
tools may
be integrated with user interface (UI) and rendering packages so that the software
receives button click events, selections, and text submissions in fields, exactly as
it would in operation.  These tools then monitor for subtle hints of problems, such
as an internal table name in an error message.
Many web app scanners include fuzzers.

Internet and web protocols require a huge amount of complex
processing that have historically been a source of serious vulnerabilities.

Penetration testing is a ``test methodology in which assessors, typically working
under specific constraints, attempt to circumvent or defeat the security features of
an information system.''~\cite{CNSSI4009-2015}  That is, it is humans using tools,
technologies, and their knowledge and expertise to simulate attackers in order to
detect vulnerabilities and exploits.

\subsubsubsection{Example Tools}

Acunetix, AppScan, AppSpider, Arachni, Burp, Contrast, Grabber, IKare, Nessus, Probely,
SQLMap, Skipfish, StackHawk, Vega, Veracode DAST, W3af, Wapiti, WebScarab, Wfuzz,
and
Zed Attack Proxy (ZAP).

\begin{table}[H]
\centering
\begin{tabular}{|p{0.65\textwidth}|p{0.25\textwidth}|}
\hline
{\bfseries Technique, Principle, or Directive} &
{\bfseries Reference} \\

\hline
``Web application scanner'' &
\cite[p.~8-5, Sec.~8.B]{SOAR2016}
\\

\hline
``Security Testing in the Test/Coding Phase'',
subsection ``System Testing'' &
% search for "penetration"
\cite{rbfstBSI2013}
\\

\hline
[ST2.1] ``Integrate black-box security tools into the QA process'' &
\cite[p.~77]{BSIMM112020}
\\

\hline
3.12.1e ``Conduct penetration testing [\emph{Assignment: organization-defined
    frequency}], leveraging automated scanning tools and ad hoc tests using subject
    matter experts.'' &
\cite{enhancedSecToProtectCUI_SP800-172_2021}
\\

\hline
\end{tabular}
\caption{Related Standards, Guides, or References for Web Application Scanning}
\end{table}

\subsection{Supplemental: Static Analysis}
\label{sec:staticAnalysis}

Static analysis or Static Application Security Testing (SAST) tools, sometimes called
``scanners'',
examine the code, either source or binary, to warn of possible weaknesses.  Use of
these tools enable early, automated problem detection.  Some tools can be accessed
from within an Integrated Development Environment (IDE), providing developers with
immediate feedback.  Scanners can find issues such as buffer overflows, SQL injections,
and
noncompliance with an organization's coding standards. The results may highlight the
precise files, line numbers, and even execution paths that are affected to aid
correction by developers.

Organizations should select and standardize on static analysis tools and establish
lists of ``must fix'' bugs based on their experience with the tool, the applications
under development, and reported vulnerabilities.  You may consult published lists of
top bugs (see Sec.~\ref{sec:noTopBugs}) to help create a process-specific list of
``must fix'' bugs.

SAST scales well as tests can be run on large software and can be run
repeatedly, as with nightly builds for the whole system or
in developers' IDE.

SAST tools have weaknesses, too.  Certain types of vulnerabilities are
difficult to find, such as authentication problems, access control issues, and
insecure use of cryptography.  In almost all tools, some warnings are false
positives, and some are insignificant in the context of the software.
Further, tools usually cannot determine if a weakness is an actual vulnerability or
is mitigated in the application.  Tool users should apply
warning suppression and prioritization mechanisms provided by tools to 
triage the tool results and focus their effort on correcting the most important
weaknesses.

Scanners have different strengths because of code styles, heuristics, and relative
importance classes of vulnerabilities have to the process.  You can realize the
maximum benefit by running more than one analyzer and paying attention to the weakness
classes for which each scanner is best.

Many analyzers allow users to write rules or patterns to increase the analyzer's
benefit.

\subsubsubsection{Example Tools}

Astr\'ee, Polyspace Bug Finder, Parasoft C/C++test, Checkmarx SAST, CodeSonar,
Co\-verity, Fortify, Frama-C, Klocwork, SonarSource, SonarQube, and Veracode SAST
handle many common compiled languages.

For JavaScript, JSLint, JSHint, PMD, and ESLint.

\begin{table}[H]
\centering
\begin{tabular}{|p{0.65\textwidth}|p{0.25\textwidth}|}
\hline
{\bfseries Technique, Principle, or Directive} &
{\bfseries Reference} \\

\hline
PW.8: Test Executable Code to Identify Vulnerabilities and Verify Compliance with
Security Requirements &
\cite{dodsonEtAlSSDF2020}
\\

\hline
``Source code quality analyzers'' & \\
``Source code weakness analyzers'' &
\cite[p.~8-5, Sec.~8.B]{SOAR2016}
\\

\hline
{[CR1.4]} ``Use automated tools along with manual review'' & \\
{[CR2.6]} ``Use automated tools with tailored rules'' &
\cite[pp.~75--76]{BSIMM112020}
\\

\hline
\end{tabular}
\caption{Related Standards, Guides, or References for Static Analysis}
\end{table}

\subsection{Supplemental: Human Reviewing for Properties}
\label{sec:propertyReview}

As discussed in Sec.~\ref{sec:staticAnalysis}, static analysis tools scan for many
properties and potential problems.  Some properties are poorly suited to computerized
recognition, and hence may warrant human examination.  This examination may be more
efficient with scans that indicate possible problems or locations of interest.

Someone other than the original author of the code may review it to ensure that it
\begin{itemize}
  \item performs bounds checks \cite{cacaBSI2008},

  \item sets initial values for data \cite{cacaBSI2008},

  \item only allows authorized users to access sensitive transactions, functions, and
    data \cite[p.~10, Sec.~3.1.2]{protectCUI_SP800-171_2020}
    (may include check that user functionality is separate from system management
    functionality~\cite[p.~37, Sec.~3.13.3]{protectCUI_SP800-171_2020}),

  \item limits unsuccessful logon attempts
    \cite[p.~12, Sec.~3.1.8]{protectCUI_SP800-171_2020},

  \item locks a session after period of inactivity
    \cite[p.~13, Sec.~3.1.10]{protectCUI_SP800-171_2020},

  \item automatically terminates session after defined conditions
    \cite[p.~13, Sec.~3.1.11]{protectCUI_SP800-171_2020},

  \item has an architecture that ``promote[s] effective information security within
    organizational systems''
    \cite[p.~36, Sec.~3.13.2]{protectCUI_SP800-171_2020},

  \item does not map memory to hardcoded locations,
    see Sec.~\ref{sec:memorySafeCompilation},

  \item encrypts sensitive data for
    transmission~\cite[p.~14, Sec.~3.1.13]{protectCUI_SP800-171_2020}
    and storage~\cite[p.~15, Sec.~3.1.19]{protectCUI_SP800-171_2020}.

  \item uses standard services and application program interfaces
    (APIs)~\cite[PW.4]{dodsonEtAlSSDF2020},

  \item has a secure default configuration~\cite[PW.9]{dodsonEtAlSSDF2020}, and

  \item has an up-to-date documented interface.

\end{itemize}

A documented interface includes the inputs, options, and configuration files.
The interface should be small, to reduce the attack
surface~\cite[p.~15]{IoTBCode2017}.

Threat modeling may indicate that certain code poses significant risks.
A focused manual review of small pieces, typically less than 100 lines, of code may
be beneficial for the cost.  The review could answer
specific questions.  For example, does the software require authorization when it
should?  Do the software interfaces check and validate inputs?

\begin{table}[H]
\centering
\begin{tabular}{|p{0.65\textwidth}|p{0.25\textwidth}|}
\hline
{\bfseries Technique, Principle, or Directive} &
{\bfseries Reference} \\

\hline
``Focused manual spot check'' &
\cite[p.~5-6, Sec.~5.A]{SOAR2016}
\\

\hline
3.14.7e ``Verify the correctness of [\emph{Assignment: organization-defined security
    critical or essential software, firmware, and hardware components}] using
    [\emph{Assignment: organization-defined verification methods or techniques}] &
\cite{enhancedSecToProtectCUI_SP800-172_2021}
\\

\hline
\end{tabular}
\caption{Related Standards, Guides, or References for Human Reviewing for Properties}
\end{table}

\subsection{Supplemental: Sources of Test Cases}
\label{sec:genSecPrincip}
\label{sec:riskAssessTests}
\label{sec:DOSandOverload}

Typically, tests are based on specifications or requirements, which are designed to
ensure the software does what it is supposed do, and process experience, which is
designed to ensure previous bugs do not
reappear.  Additional test cases may be based on the following principles:
\begin{itemize}
    \item Threat modeling---concentrate on areas with highest consequences,
    \item General security principles---find security vulnerabilities, such as
    failure to check credentials, since these often do not cause
    operational failures (crashes or incorrect output),
    \item Negative tests---make sure that software behaves reasonably for
    invalid inputs and that it does not do what it should
    \emph{not} do, for instance ensure that a user cannot perform operations for
    which they are not authorized~\cite[p.~8-5, Sec.~8.B]{SOAR2016},
    \item Combinatorial testing---find errors occurring when
    handling certain n-tuples of kinds of 
    input~\cite{combinTestKuhnetal2009, combinTestKuhnetal2015}, and
    \item Denial of service and overload---make sure software is resilient.
\end{itemize}

Denial of service and overload tests are also called stress testing.
Also consider algorithmic attacks.
Algorithms may work well with a typical load or expected overload, but an attacker 
may cause a load many orders of magnitude higher than would ever occur in actual
use.

Monitor execution and output during negative testing especially,
such as with Interactive Application Security Testing (IAST) tools or
Runtime Application Self-Protection (RASP).

\begin{table}[H]
\centering
\begin{tabular}{|p{0.65\textwidth}|p{0.25\textwidth}|}
\hline
{\bfseries Technique, Principle, or Directive} &
{\bfseries Reference} \\

\hline
``Simple attack modeling'' & \\
``Negative testing'' &
\cite[pp.~8-4 and 8-5, Sec.~8.B]{SOAR2016}
\\

\hline
%      General security principles
``Security Testing in the Test/Coding Phase'',
subsection "Unit Testing" and
% search for "security principals" <<< TYPO ON WEB SITE
%``Security Testing in the Test/Coding Phase'',
subsection ``System Testing'' & \\
% search for "penetration"
``Security Testing Activities'',
subsection ``Risk Analysis'' &
% search for "select test data inputs"
\cite{rbfstBSI2013}
\\

\hline
{[AM1.2]} ``Create a data classification scheme and inventory'' & \\
{[AM1.3]} ``Identify potential attackers'' & \\
{[AM2.1]} ``Build attack patterns and abuse cases tied to potential attackers'' & \\
{[AM2.2]} ``Create technology-specific attack patterns'' & \\
{[AM2.5]} ``Build and maintain a top N possible attacks list'' & \\
{[AM3.2]} ``Create and use automation to mimic attackers'' &
\cite[pp.~67--68]{BSIMM112020}
\\

\hline
{[ST1.1]} ``Ensure QA performs edge/boundary value condition testing'' & \\
{[ST1.3]} ``Drive tests with security requirements and security features'' & \\
{[ST3.3]} ``Drive tests with risk analysis results'' &
\cite[pp.~77--78]{BSIMM112020}
\\

% IAST and RASP
\hline
{[SE1.1]} ``Use application input monitoring'' & \\
{[SE3.3]} ``Use application behavior monitoring and diagnostics'' &
\cite[pp.~80 and 82]{BSIMM112020}
\\

\hline
3.11.1e Employ [\emph{Assignment: organization-defined sources of threat
    intelligence}] as part of a risk assessment to guide and inform the development
of organizational systems, security architectures, selection of security solutions,
monitoring, threat hunting, and response and recovery activities &
%\cite{enhancedSecToProtectCUI_SP800-172_2021}
\\

3.11.4e Document or reference in the system security plan the security solution
selected, the rationale for the security solution, and the risk determination &
\cite{enhancedSecToProtectCUI_SP800-172_2021}
\\

\hline
\end{tabular}
\caption{Related Standards, Guides, or References for Sources of Test Cases}
\end{table}

\subsection{Supplemental: Top Bugs}
\label{sec:noTopBugs}

There are many collections of high priority bugs and weaknesses, such as those
identified in the Common Weakness Enumeration (CWE)/SANS Top 25 Most
Dangerous Software Errors~\cite{aboutCWE, CWESANSTop25},
the CWE Weaknesses on the Cusp~\cite{CWETop25}, or 
the Open Web Application Security Project (OWASP) Top 10
Web Application Security Risks~\cite{OWASPTop10}.

These lists, along with experience with bugs found, can help developers begin
choosing bug classes to focus on during
verification and process improvement.

\begin{table}[H]
\centering
\begin{tabular}{|p{0.65\textwidth}|p{0.25\textwidth}|}
\hline
{\bfseries Technique, Principle, or Directive} &
{\bfseries Reference} \\

\hline
``UL and Cybersecurity'' &
\cite[slide 8]{ULandCybersecurity2011}
\\

\hline
For security ``Code Quality Rules''
lists 36 ``parent'' CWEs and 38 ``child'' CWEs.
For reliability, it lists 35 ``parent'' CWEs and 39 ``child'' CWEs. &
\cite{CISQQualRules}
\cite{CISQListOfWeaknesses2019}
\\

\hline
\end{tabular}
\caption{Related Standards, Guides, or References for Top Bugs}
\end{table}

\subsection{Supplemental: Checking Included Software for Known Vulnerabilities}
\label{sec:noKnownVulns}

You need to have as much assurance for included code, e.g., closed source software,
free and open-source software, libraries, and packages, as for code you develop.  If
you lack strong guarantees, we recommend as much testing of included code as of the
code.

Earlier versions of packages and libraries may have known vulnerabilities
that are corrected in later versions.

Software Composition Analysis (SCA) and Origin Analyzer (OA) tools scan a code base
to identify what code is included.  They also check for any vulnerabilities that have
been reported for the included code~\cite[App. C.21, p.~C-44]{SOAR2016}.
A widely-used database of publicly known vulnerabilities is the NIST National
Vulnerability Database (NVD), which identifies vulnerabilities using
Common Vulnerabilities and Exposures (CVE).
Some tools can be configured to prevent download of software that has security
issues and recommend alternative downloads.

Since libraries are matched against a tool's database, they will not identify
libraries missing from the database.

\subsubsubsection{Example Tools}

Black Duck, Binary Analysis Tool (BAT), Contrast Assess, FlexNet Code Insight, FOSSA,
JFrog Xray, OWASP Dependency-Check, Snyk, Sonatype IQ Server, Veracode SCA, WhiteHat
Sentinel SCA, and WhiteSource Bolt.

\begin{table}[H]
\centering
\begin{tabular}{|p{0.65\textwidth}|p{0.25\textwidth}|}
\hline
{\bfseries Technique, Principle, or Directive} &
{\bfseries Reference} \\

\hline
``Origin Analyzer'' &
\cite[App. C.21, p.~C-44]{SOAR2016}
\\

\hline
``UL and Cybersecurity'' &
\cite[slide 8]{ULandCybersecurity2011}
\\

\hline
3.4.3e Employ automated discovery and management tools to maintain an up-to-date,
complete, accurate, and readily available inventory of system components &
\cite{enhancedSecToProtectCUI_SP800-172_2021}
\\

\hline
\end{tabular}
\caption{Related Standards, Guides, or References for Checking Included Software for Known Vulnerabilities}
\end{table}

\section{Beyond Software Verification}
\label{sec:beyondMinimum}

Good software must be built well from the beginning.
Verification is just one element in delivering
software that meets operational security 
requirements.  The software assurance techniques listed
above are just the minimum steps to use in improving the
security of enterprise supply chains.
Section~\ref{sec:goodDevelopment} describes a 
few general software development practices and how assurance fits
into the larger subject of secure software development and 
operation.  Even software that has solid security characteristics 
can be exploited by adversaries if its 
installation, operation, or maintenance is conducted in a manner 
that introduces vulnerabilities.
Section~\ref{sec:goodOperation} describes some trends and
technologies that may improve software assurance.
Section~\ref{sec:betterSWAssurance} describes good 
installation and operation principles.  Both software 
development and security technologies are constantly evolving. 

\subsection{Good Software Development Practices}
\label{sec:goodDevelopment}

Ideally, software is secure by design, and the security of both 
the design and its implementation can be demonstrated, 
documented, and maintained.  Software development, and indeed the 
full software development lifecycle, has changed over time, but 
some basic principles apply in all cases.  NIST developed a cybersecurity
white paper, ``Mitigating the Risk of Software Vulnerabilities by Adopting a Secure
Software Development Framework (SSDF)''~\cite{dodsonEtAlSSDF2020},
that provides an overview and references
about these basic principles.  This document is part of an ongoing project; see
\url{https://csrc.nist.gov/Projects/ssdf}.  The SSDF introduces a software
development
framework of fundamental, sound, and secure software development practices based on
established secure software development practice documents.  For verification to
be most effective, it should be a part of the larger software development
process.  SSDF practices are organized into four groups:
\begin{itemize}
  \item Prepare the Organization (PO): Ensure that the organization's people,
    processes, and technology are prepared at the organization level and, in some
    cases, for each individual project to develop secure software.
  \item Protect the Software (PS): Protect all components of the software from
    tampering and unauthorized access.
  \item Produce Well-Secured Software (PW): Produce well-secured software that has
    minimal security vulnerabilities in its releases.
  \item Respond to Vulnerabilities (RV): Identify vulnerabilities in software
    releases, respond appropriately to address those vulnerabilities, and prevent
    similar vulnerabilities from occurring in the future.
\end{itemize}

In the context of DevOps, enterprises with secure development include the following
characteristics:
\begin{itemize}
    \item The enterprise creates a culture where security is everyone's
      responsibility.  This includes integrating a security specialist into the
      development team, training all developers to know how to design and implement
      secure software, and using automated tools that
      allow both developers and security staff to track vulnerabilities.
    \item The enterprise uses tools to automate security checking, often referred to
      as Security as Code~\cite{BoyerSecAsCode2018}.
    \item The enterprise tracks threats and vulnerabilities, in addition to typical
      system metrics.
    \item The enterprise shares software development task information, security
      threat, and vulnerability knowledge between the security team, developers,
      and operations personnel.
\end{itemize}

\subsection{Good Software Installation and Operation Practices}
\label{sec:goodOperation}

As stated above, even software that has no identified security
vulnerabilities can be subject to exploitation by adversaries if
its installation, operation, or maintenance introduces vulnerabilities.
Some issues that are not directly addressed in this paper include
misconfiguration, violation of file permission policies, network 
configuration violations, and acceptance of counterfeit or 
altered software.
See especially ``Security Measures for ``EO-Critical Software'' Use Under Executive
Order (EO) 14028'',
\url{https://www.nist.gov/itl/executive-order-improving-nations-cybersecurity/security-measures-eo-critical-software-use-under},
which addresses patch management, configuration
management, and continuous monitoring, among other security measures, and also
provides a list of
references.

\underline{Configuration files}: Because of the differences in software
applications and networking environments, the parameters 
and initial settings for many computer applications, server 
processes, and operating systems are configurable.
Often, security verification fails to anticipate unexpected
settings.  Systems and network operators often alter 
settings to facilitate tasks that are more
difficult or infeasible when using restrictive settings. 
Particularly in cases of access authorization and network 
interfaces, changing configuration settings can introduce 
critical vulnerabilities.  Software releases should include secure
default settings and caveats regarding deviations from 
those settings.  Security verification should include all valid
settings and (possibly) assurance that invalid settings 
will be caught by run-time checks.  The acquirer should be
warned or notified that settings other than those explicitly 
permitted will invalidate developer's security assertions.

\underline{File Permissions}: File ownership and permissions to read,
write, execute, and delete files need to be established
using the principle of least privilege.  No matter how
thoroughly software has been verified, security is compromised if it
can be modified or if files can be accessed by unauthorized 
entities.  The ability to change file permissions needs to 
be restricted to explicitly authorized subjects that are 
authenticated in a manner that is commensurate with the 
impact of a compromise of the software.  The role of file 
permissions in maintaining security
assertions needs to be explicit.

\underline{Network configuration}: Secure configuration refers to
security measures that are implemented when building and 
installing computers and network devices to reduce
cyber vulnerabilities.  Just as file permissions 
are critical to the continued integrity of software, so 
does network configuration constrain unauthorized access to 
software.  Verification needs to cover all valid network
configuration settings and (possibly) provide assurance 
that invalid settings will be caught by run-time checks. 
The role of network configuration in scoping the 
applicability of security assertions needs to be explicit.

\underline{Operational configuration}: Software is employed in a
context of use.  Addition or deletion of components that are 
dependent on a software product or on which the product 
depends can either validate or invalidate the assumptions 
on which the security of software and system operation
depend.  Particularly in the case of source code, the 
operating code itself depends on components such as 
compilers and interpreters.  In such cases, the security of 
the software can be invalidated by the other products. 
Verification needs to be conducted in an environment that is
consistent with the anticipated operational configurations. 
Any dependence of the security assertions on implementing 
software or other aspects of operational configuration 
needs to be made explicit by the developer.
Supply chain integrity must be maintained.

\subsection{Additional Software Assurance Technology}
\label{sec:betterSWAssurance}

Software verification continues to improve as new methods are developed
and existing methods
are adapted to changing development and operational environments.
Some challenges remain, e.g., applying formal methods to prove the correctness of
poorly designed code.
Nearer-term advances that may add to security assurance based on
verification include:
\begin{itemize}
    \item Applying machine learning to reduce false positives from automated
    security scanning tools and to increase the vulnerabilities that these tools
    can detect.
    \item Adapting tools designed for automated web interface tests, e.g., Selenium,
    to produce security tests for applications.
    \item Improving scalability of model-based security testing for complex systems.
    \item Improving automated web-application security assessment tools with respect to:
    \begin{itemize}
        \item Session state management
        \item Script parsing
        \item Logical flow
        \item Custom uniform resource locators (URLs)
        \item Privilege escalation
    \end{itemize}
    \item Applying observability tools to provide security assurance in cloud environments.
    \item Adapting current security testing to achieve cloud service security assurance.
\end{itemize}

Other techniques to reduce software vulnerabilities are described in ``Dramatically
Reducing Software Vulnerabilities'', NIST-IR 8151~\cite{Black2016DReduceSwVuls}.

\section{Documents Examined}
\label{sec:sourceDocs}

This section lists some of the standards, guides, references, etc., that we examined
to assemble this document.  We list them to give future work an
idea where to start or
quickly learn what may have been overlooked.  We group related references.

\vspace{2ex}

Donna Dodson, Murugiah Souppaya, and Karen Scarfone, ``Mitigating the Risk of
Software Vulnerabilities by Adopting a Secure Software Development Framework
(SSDF)'', 2013 \cite{dodsonEtAlSSDF2020}.

\vspace{2ex}

David A. Wheeler and Amy E. Henninger,
``State-of-the-Art Resources (SOAR) for Software Vulnerability Detection, Test, and
Evaluation 2016'', 2016
\cite{SOAR2016}.

\vspace{2ex}

Steven Lavenhar,
``Code Analysis'', 2008
\cite{cacaBSI2008}.

C.C. Michael, Ken van Wyk, and Will Radosevich,
``Risk-Based and Functional Security Testing'', 2013
\cite{rbfstBSI2013}.

\vspace{2ex}

UL, ``IoT Security Top 20 Design Principles'', 2017
\cite{ULIoTSecPrinciples2017}.

\vspace{2ex}

Tom Haigh and Carl E. Landwehr, ``Building Code for Medical Device Software
Security'', 2015 \cite{medicalDevBCode2015}.

Ulf Lindqvist and Michael Locasto, ``Building Code for the Internet of Things'', 2017
\cite{IoTBCode2017}.

Carl E. Landwehr and Alfonso Valdes, ``Building Code for Power System Software
Security'', 2017 \cite{powerSysBCode2017}.

\vspace{2ex}

``Protection Profile for Application Software Version 1.3'', 2019
\cite{PPforAppSoftware2019v1.3}.

\vspace{2ex}

Ron Ross, Victoria Pillitteri, Gary Guissanie, Ryan Wagner, Richard Graubart, and Deb
Bodeau, ``Enhanced Security Requirements for Protecting Controlled Unclassified
Information: A Supplement to NIST Special Publication 800-171'', 2021
\cite{enhancedSecToProtectCUI_SP800-172_2021}.

Ron Ross, Victoria Pillitteri, Kelley Dempsey, Mark Riddle, and Gary Guissanie,
``Protecting Controlled Unclassified Information in Nonfederal Systems and
Organizations'', 2020
\cite{protectCUI_SP800-171_2020}.

Ron Ross, Victoria Pillitteri, and Kelley Dempsey, ``Assessing Enhanced Security
Requirements for Controlled Unclassified Information'', 2021
\cite{AssesEnhancedSecForCUI_SP800-172A_2021}.

\vspace{2ex}

Bill Curtis, Bill Dickenson, and Chris Kinsey, ``CISQ Recommendation Guide: Effective
Software Quality Metrics for ADM Service Level Agreements'', 2015
\cite{CISQSWQualMetrics2015}.

``Coding Quality Rules'', 2021
\cite{CISQQualRules}.

\section{Glossary and Acronyms}

\begin{table}[H]
\centering
\begin{tabular}{p{0.3\textwidth}p{0.65\textwidth}}
%\hline
{\bfseries Term} &
{\bfseries Definition} \\

Cybersecurity &
The practice of protecting systems, networks, and programs from digital attacks. \\

Software Source Code &
The software as it is originally entered
in plain text, e.g., human-readable alphanumeric characters. \\

\end{tabular}
\label{tab:glossary}
\end{table}

\begin{table}[H]
\centering
\begin{tabular}{p{0.1\textwidth}p{0.85\textwidth}}

API & Application Program Interface \\
CVE & Common Vulnerabilities and Exposures \\
CWE & Common Weakness Enumeration \\
DAST & Dynamic Application Security Testing \\
EO & Executive Order \\
HW & Hardware \\
IAST & Interactive Application Security Testing \\
MISRA & Motor Industry Software Reliability Association \\
NIST & National Institute of Standards and Technology \\
NSA & National Security Agency \\
NVD & National Vulnerability Database \\
OA & Origin Analyzer \\
OS & Operating System \\
OSS & Open Source Software \\
OWASP & Open Web Application Security Project \\
RASP & Runtime Application Self-Protection \\
SAST & Static Application Security Testing \\
SCA & Software Composition Analysis \\
SDLC & Software Development Life Cycle \\
SSDF & Secure Software Development Framework \\
URL & Uniform Resource Locator \\

\end{tabular}
\label{tab:acronyms}
\end{table}

\section*{References}
\addcontentsline{toc}{section}{References}
\bibliographystyle{techpubs}
\bibliography{references}

%%%%%%%%%%%%%%%%%%%%%%%%%%%%%%%%%%%%%%%%%%%%%%%%%%%%%%%%%%%%%%%%%%%%
%   Please use the techpubs BibTeX style when compiling bibliography, or follow
%   the instructions on tinyurl.com/techpubsnist to format your .bib / .bbl file
%   appropriately.
%%%%%%%%%%%%%%%%%%%%%%%%%%%%%%%%%%%%%%%%%%%%%%%%%%%%%%%%%%%%%%%%%%%%

\end{document}